\documentclass[sigconf]{acmart}
\usepackage{algorithm}
\usepackage{algorithmic}
\usepackage{amsmath}
\usepackage{adjustbox}

\AtBeginDocument{%
  \providecommand\BibTeX{{%
    \normalfont B\kern-0.5em{\scshape i\kern-0.25em b}\kern-0.8em\TeX}}}

\setcopyright{acmcopyright}
\copyrightyear{2021}
\acmYear{2021}
\acmDOI{}

\acmConference[]{ICAIF}{Workshop on Explainable AI in Finance (XAIF)}{2021}
\acmBooktitle{ICAIF, Workshop on Explainable AI in Finance (XAIF)}
\acmPrice{}
\acmISBN{}

\begin{document}

\title{Parameterized Explanations for Investor / Company Matching}

\author{Simerjot Kaur}
\email{simerjot.kaur@jpmchase.com}
\affiliation{%
  \institution{J.P.Morgan AI Research}
  \city{Palo Alto}
  \state{California}
  \country{USA}
}

\author{Ivan Brugere}
\email{ivan.brugere@jpmchase.com}
\affiliation{%
  \institution{J.P.Morgan AI Research}
  \city{Palo Alto}
  \state{California}
  \country{USA}
}

\author{Andrea Stefanucci}
\email{andrea.stefanucci@jpmorgan.com}
\affiliation{%
  \institution{J.P.Morgan AI Research}
  \city{New York}
  \state{NY}
  \country{USA}
}

\author{Armineh Nourbakhsh}
\email{armineh.nourbakhsh@jpmchase.com}
\affiliation{%
  \institution{J.P.Morgan AI Research}
  \city{New York}
  \state{NY}
  \country{USA}
}

\author{Sameena Shah}
\email{sameena.shah@jpmchase.com}
\affiliation{%
  \institution{J.P.Morgan AI Research}
  \city{New York}
  \state{NY}
  \country{USA}
}

\author{Manuela Veloso}
\email{manuela.veloso@jpmchase.com}
\affiliation{%
  \institution{J.P.Morgan AI Research}
  \city{New York}
  \state{NY}
  \country{USA}
}

\renewcommand{\shortauthors}{Kaur, et al.}

\begin{abstract}
  Matching companies and investors is usually considered a highly specialized decision making process. Building an AI agent that can automate such recommendation process can significantly help reduce costs, and eliminate human biases and errors. However, limited sample size of financial data-sets and the need for not only good recommendations, but also explaining why a particular recommendation is being made, makes this a challenging problem. In this work we propose a representation learning based recommendation engine that works extremely well with small datasets and demonstrate how it can be coupled with a parameterized explanation generation engine to build an explainable recommendation system for investor-company matching. We compare the performance of our system with human generated recommendations and demonstrate the ability of our algorithm to perform extremely well on this task. We also highlight how explainability helps with real-life adoption of our system.  
\end{abstract}

% %%
% %% The code below is generated by the tool at http://dl.acm.org/ccs.cfm.
% %% Please copy and paste the code instead of the example below.




\maketitle

\section{Introduction}\label{intro}
Matching companies to investors, and investors to companies has long been considered a highly specialized human cognitive decision making process. Such match-making exercise is extremely beneficial to both investors and startups/companies who are looking for funding. On one end, it helps investors identify which companies to invest in, and on the other end, it helps provide guidance to startup/companies on which investors to approach when they are looking to raise funding. Moreover, in financial organizations, this match-making can have direct financial consequences, for both investors and companies. Hence the ability to explain, why a particular investment opportunity is beneficial to investors and why a startup should approach a particular investor for funding, becomes critical.

However, matching investor and companies and providing explanations is typically very expensive as it involves manual curation of data about companies and investors, analyzing their historical track records, performance, and preferences, hence becoming a tedious task involving significant amounts of manual labour. Further, it brings about human errors and the results are often dependent on the individual experience and expertise of humans involved in this process. 

To overcome the above limitations of high cost, and need for a human expert with years of experience, we explore the possibility of matching companies and investors in an automated manner by building a deep-learning based investor-company matching recommendation engine. However, unlike most e-commerce recommendation engines, such as netflix movie recommendations, recommendation systems for financial applications need to be trained on extremely small and limited datasets, and they should not only provide good recommendations, but must also be capable of providing explanations as to why a particular recommendation is being made. This is critical to ensure real-life adoption of such models. In-fact, the limited size of dataset and lack of explainability of current deep learning models poses a serious challenge to adoption within the financial domain. In this work we propose a novel algorithm that leverages recent advances in representation learning to build a recommendation engine under small data-sets, and combines it with a parameterized explanation generation scheme to build an explainable investor-company matching system.

\begin{figure}[hb]
\begin{center}
\centerline{\includegraphics[width=\columnwidth]{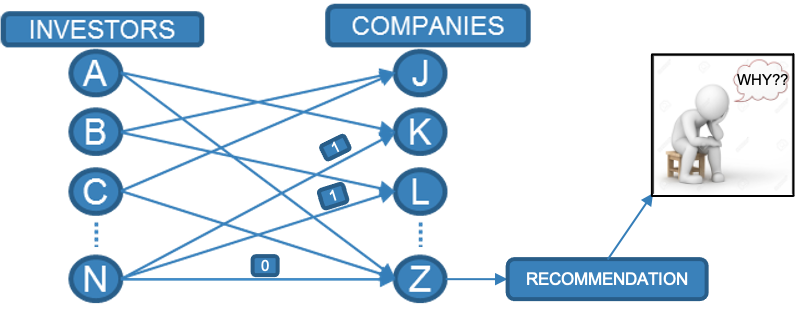}}
\caption{Problem Statement}
\label{fig:Intro}
\end{center}
\vskip -0.2in
\end{figure}

Formally, figure \ref{fig:Intro} above summarizes our problem. \it Given a limited sample-set of investors ${A,B...N}$, and companies ${J,K,...Z}$, together with investor/company attributes and their historical interactions, can we recommend new links, i.e. recommend new companies to investors, and investors to companies while also providing an explanation for why a particular recommendation has been made? \rm.  

Our proposed approach consists of first building a high dimensional vector representation for each investor and company using available input data such as their description, their past investments etc. and subsequently using these vector representations as inputs to our hybrid content and collaborative model, to generate similarity scores between investors and companies which are used for making recommendations. Further, we utilize the outputs generated by our content and collaborative based model to extract parameters for a parameterized template based explanatory engine. We demonstrate that our algorithm can perform extremely well on investor-company matching task while also alleviating human biases and errors. Further, we also highlight how our parameterized explanation algorithm helps improve adoption of our algorithm within financial institutions. We also verified that our proposed approach is fair and it is not biased based on management team's personal profile or location of company/investor.

The rest of the paper is outlined as follows. Section \ref{relwork} describes related work and highlights the unique challenges that come about in building an investor-company matching recommendation engine within financial domain, section \ref{data} describes the details of our dataset and section \ref{matching} provides details of our representation learning approach and describes our proposed hybrid recommendation system in more details. Section \ref{algo} provide the details of our parameterized explanatory algorithm, and demonstrates the ability of our approach to algorithmically generate explanations for why a particular company-investor pair could be linked. Finally, section \ref{conclusion} concludes our work and lays out foundation for future work in this area. 
 
\vspace*{-7mm}

\section{Related Work}\label{relwork}

Various versions of deep learning based recommendation systems have been developed by major organizations such as Facebook \cite{naumov2019deep}, Netflix \cite{GomezUribe2015TheNR}, Amazon \cite{Linden2003AmazoncomRI}, etc. While many of these e-commerce based recommendation system algorithms have achieved remarkable success, they are almost always built using some form of supervised learning approaches. This is in-part due to the easy availability of labeled data. For example, it is relatively easy to gather movie/item ratings from users, allowing them to use supervised learning approaches. However, such labeled datasets do not exist when we are trying to match companies and investors. In-fact, the only interaction information amongst the investors and companies that we have access to is a binary link informing whether an investor  $i$ has invested in a company $j$ or not. We do not have any information on how much an investor likes or dislikes a company. Even if they liked a company, they might still end up not investing in it because of other factors. Hence, unlike many e-commerce recommendation systems our system is unsupervised in nature, making it difficult to directly apply existing methods. Further, as explained in the introduction section, unlike e-commerce application, adoption in financial applications is heavily dependent on our ability to provide explanations for why we are recommending an investor to a company or vice-versa. Such explanations are rather limited in many existing deep learning based approaches. 

Additionally, while there has been some work in the literature on trying to match investors for startups \cite{IJCCC}, \cite{XU2020103}, \cite{pub.1137411524}, \cite{10.1145/2505515.2507882}, \cite{10.1145/2792838.2800181}, almost all these approaches are largely based on statistical techniques and do not leverage recent advances in deep learning, which has been the driver of improvement in recommendation engines over the past few years. Further, most of this work also overlooks the essential explainability requirements that are essential for financial applications. 

In-fact, to the best of our knowledge our work here is the first end-to-end investor-company matching system that leverages advances in deep learning and unsupervised learning while also coupling it with a parameterized explanation generation engine, helping to improve adoption of deep learning into financial applications. 

\vspace*{-2mm}

\section{Data}\label{data}

This section describes the details for our dataset. Our model has been developed and tested using licensed data. Essentially, the dataset used contains various details of startup companies such as a brief description about what the company does, its industry focus, information about its previous deals such as the series raised, and amount raised in that series etc. Similarly, for investors we have certain characteristic information about the investors such as their funding style and industry preference. Finally, we also have a historical binary investor-company link matrix. 

We further describe the features we use from the dataset in more details below : 

\begin{itemize}
  \item{\bf Company Data}: The company features include its description, industry focus, year founded, location, etc. For instance, some of the information on company \it ABC Lane \rm \footnote{Due to proprietary reasons, the company and investor names have been masked and fictitious names have been used.} include:
  \begin{itemize}
    \item{\it Description}: Operator of a consumer finance intended to help consumers have access to fair and clear credit. The company leverages advanced technology, data analytics, and machine learning to provide a dignified customer experience to people who are working hard to build or rebuild their credit and have terms that are better and easier to understand than most of the alternatives available to people with less-than-pristine credit or limited credit history, enabling consumers to access credit.
    \item{\it Industry Focus}: Financial Services/Other Financial Services/Consumer Finance/Financial Software
    \item{\it Year Founded}: 2018
    \item{\it Location}: Atlanta, Georgia, United States
  \end{itemize}
  \item{\bf Deals}: The details regarding the capital raises done by the company include the type of series rounds that the company has gone through, when were the deals made, the amount of capital raised in each round, etc. For instance, some of the information on the deals done by the company \it ABC Lane \rm include:
    \begin{itemize}
    \item{\it Types of  Series Rounds}: Seed round, Series A round
    \item{\it Amount Raised}: \char`\~\$50 million
    \item{\it Deal Dates}: year 2018 and 2019
  \end{itemize}
  \item{\bf Investor Data}: The investor features has been derived from the dataset including its funding preferences, industry preferences, location, etc. For instance, features on one of the investors \it The ABC Group \rm of company \it ABC Lane \rm include:
\begin{itemize}
    \item{\it Funding Style}: Mostly involved in initial stage of funding, specifically, invested 27\%, 23\% and 20\% of the time in Seed, Series A and Series B rounds respectively and the remaining in later stages of capital raise. 
    \item{\it Industry Preferences}: Mostly involved in healthcare and consumer products and services, specifically, invested 56\% and 20\% of the time in Healthcare and Consumer Products and Services respectively and the remaining in other sectors like Information Technology, Business Products and Services, etc.
    \item{\it Location}: New York, NY, United States
  \end{itemize}
  \item{\bf Historical Investor-Company Link Matrix}: This binary link matrix contains historical information on whether an investor $i$ has invested in a company $j$ or not.
\end{itemize}

\noindent In order to test the investor-company link matching algorithm, we split the historical investor-company link matrix into 70\% training and 30\% test set. Given that our model is unsupervised in nature, we create two types of test sets : 
\begin{itemize}
\item{\bf Test Set containing Historical Links}: This set contains a list of investor and company pairs for which there is historical information that a link exist between the investors and companies i.e. the investors have invested in the companies historically. In this case, we would expect an ideal model to predict 100\% links between the investors and companies. 
\item{\bf Test Set without Historical Links}: This set contains a set of randomly chosen pair of investor and companies for which there is information that there does not exist a link between the investors and companies i.e. the investors have not invested in these companies in the past. In this case, we expect the model to predict NO links between the investors and companies for majority of the time and predict that link might exist for some percentage of the dataset which would then become our recommendations to the investors or companies. 
\end{itemize}

\section{Investor/Company Matching}\label{matching}

This section describes our algorithm used to match the companies to investors and vice versa based on each their historical track records and preferences. Subsection \ref{results} demonstrates the performance of the matching algorithm on our two test sets described in section \ref{data} above. Finally, subsection \ref{analysis} details the results obtained during human analysis, during which some of the top recommendations produced by the algorithm were given to the subject matter experts for review.

\subsection{Algorithm}

\begin{figure}[ht!]
\begin{center}
\centerline{\includegraphics[width=\columnwidth]{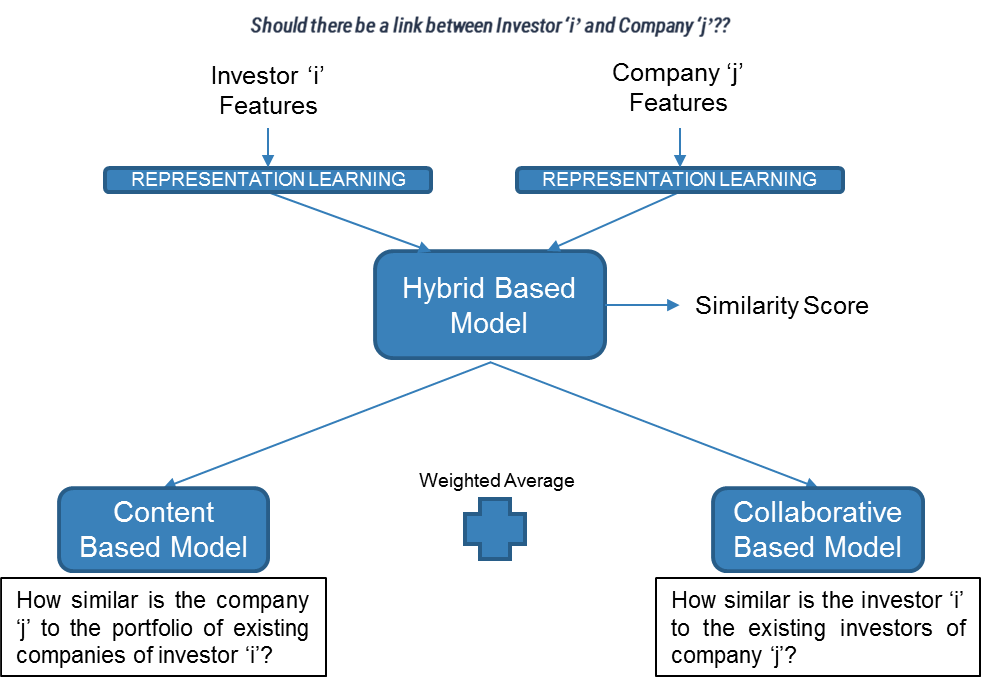}}
\caption{Proposed Approach}
\label{fig:Proposed Approach}
\end{center}
\vskip -0.3in
\end{figure}

Figure \ref{fig:Proposed Approach} describes our proposed approach to solve the investor-company matching problem, i.e.,  \it 'Should there be a link between an Investor i and Company j?' \rm. Our proposed approach consists of two sequential steps. In the first step we learn a high dimensional vector representation for each company and investor from the given unstructured text that defines the attributes/features of the company/investors. In the second step we use these vector representations to calculate the similarity between companies and investors using a hybrid approach that calculates a weighted average of similarity scores obtained from content based and collaborative based approaches. We describe each of the parts in more details in the following sections. 

\begin{figure*}[hbt!]
\begin{center}
\centerline{\includegraphics[width=1.0\textwidth]{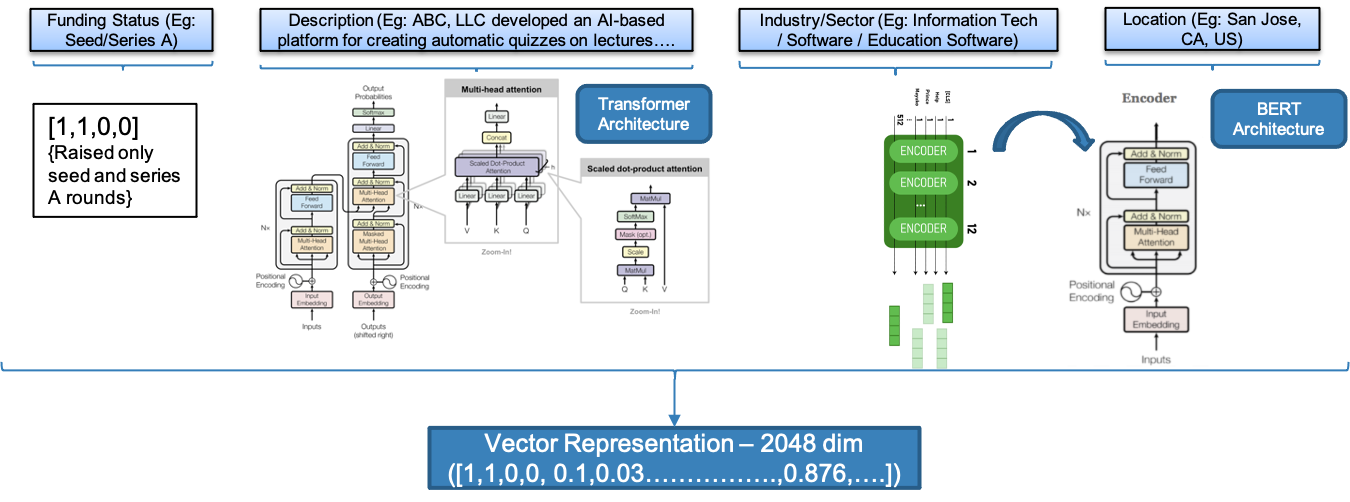}}
\caption{Company/Investor Representation: Transformer and BERT architecture graphics are excerpted from \cite{vaswani2017attention} and \cite{devlin2019bert}}
\label{fig:Representation}
\end{center}
\vskip -0.2in
\end{figure*}

\subsubsection{Representation Learning}
\hfill\\

As discussed in section \ref{data} each investor and company is characterized by multiple attributes, many of which are sequences of unstructured text. To enable us to easily perform comparisons and estimate how close/far various companies and investors are, we first learn a high-dimensional vector representation for each company and investor. The key guiding principle behind our structured high dimensional vector representation is that similar sentences/words, or phrases within a similar contexts should map close to each other in a high-dimensional space, and sentences/words or phrases that are very different should be far away from each other.

Figure \ref{fig:Representation} depicts in more detail how different kinds of unstructured text are encoded in a high dimensional distributed vector representation to generate one final representation for each company/investor. More specifically,
 \begin{itemize}
     \item{\bf Company Representation}: We represent a company by its funding status, description, industry focus, and location. The methods used to build a vector representation for each attribute are as follows:
     \begin{itemize}
         \item {\it Funding Status}: This attribute describes in which types of round of raises the company has been involved in. This feature is being represented as a Multi-Label Binarizer. For instance, if a company has only been involved in Seed and Series A rounds in the set of {Seed Round, Series A, Series B, Series C}, this feature will be represented as [1,1,0,0].
         \item {\it Description}: This attribute describes the business that the company is involved in. Since the position of the words in the description is important, pretrained transformer with self attention \cite{vaswani2017attention} based architecture has been leveraged to encode the description into a 768-dimensional dense vector representation.  
         \item {\it Industry Focus}: This attribute describes the type of industry the company operates in. Since the position of words is not crucial, pretrained BERT \cite{devlin2019bert} based architecture has been used to create 768-dimensional vector embeddings.
         \item {\it Location}: This attribute describes the city, state and country the company is located in. Similar to industry focus attribute, pretrained BERT based architecture has been used to create 768-dimensional vector embeddings.
     \end{itemize}
     Finally, the vector representations of the above attributes are concatenated together to form a full vector representation for the company.
     \item{\bf Investor Representation}: The attributes that have been used to represent a investor are its funding style, industry preference and location. The methods used to build a vector representation for each attribute are as follows:
     \begin{itemize}
         \item {\it Funding Style}: This attribute describes the amount of deals that the investor has done in each type of round raise.  For instance, if an investor has invested 80, 10 and 10 times in Seed, Series A and Series B rounds in the set of {Seed Round, Series A, Series B, Series C}, then this feature will be represented as [0.8, 0.1, 0.1, 0].
         \item {\it Industry Preference}: This attribute describes the amount of deals that the investor has done in various industry sectors. In order to build the representation for this attribute, as described in company representation, the industry sectors are first encoded using pretrained BERT based architecture and then the weighted average of these representations are obtained with weights corresponding to number of deals that the investor has done in each industry sector.
         \item {\it Location}: This attribute describes the city, state and country the investor is located in. Similar to company representation, pretrained BERT based architecture has been used to create 768-dimensional vector embeddings.
     \end{itemize}
     Finally, the vector representations of the above attributes are concatenated together to form a full vector representation for the investor.
 \end{itemize}
 
\subsubsection{Similarity Score Estimation}
\hfill\\

Given the high dimensional distributed representations for both companies and investors, we can easily estimate the closeness or remoteness between pairs of companies and/or investors following a hybrid approach, i.e. a weighted average of content  and collaborative based similarity scores. Specifically, our proposed similarity estimation method consists of following key steps : 

%\hfill\\
(A) \emph{Content-based Model:} Content-based approach uses the attributes of the companies to recommend similar companies to an investor. It recommends a company to an investor based on its existing portfolio of companies. This method relies only on company features and not on investor's preferences. Formally, for a company $j$ and investor $i$, we are trying to estimate which company and with what score $j$ is closest to within the portfolio of companies in which $i$ has invested. 

%\hfill\\
(B) \emph{Collaborative Based Model:} Collaborative-based approach generates recommendations by deriving investor's historical preferences on companies.  It recommends investors to a company based on its existing portfolio of investors. This method relies only on the historical interaction between the investors and companies as well as historical preferences of the investors and not on company features. Formally, for an investor $i$ and company $j$, we need to determine which investor and with what score $i$ is closest to within the portfolio of $j$'s previous investors.

%\hfill\\
It may be noted that since the investor-company interaction matrix is very sparse, using only one of the above similarity scores leads to performance degradation. To overcome this we obtain the final score as a weighted average score from the results of A and B above. Specifically, combining both the models helps overcome the performance degradation caused by the sparse historical interaction between the investors and companies, and it also helps alleviate cold-start problems when investors have no or very few interactions or for new companies who have not raised any funding round or interacted with any investors.

\noindent The \it Algorithm \ref{alg1} \rm describes our overall matching algorithm in a step-by-step manner. 

\begin{algorithm}
 \caption{Similarity Score Algorithm}
 \label{alg1}
 \begin{algorithmic}[1]
 \STATE \textbf{Input 1:} For an investor $i$ where $i\in [1, 2, 3, \hdots, n]$ where $n$ is the number of investors in the dataset
 \STATE \textbf{Input 2:} For a company $j$ where $j\in [1, 2, 3, \hdots, m]$ where $m$ is the number of companies in the dataset
 \STATE \textbf{Initialization:} Set content based score, $CBS$, $=0$, closest company, $CC$, $=""$, closest company business, $CCB$, $=""$ collaborative based score, $CB$, $=0$, closest investor, $CI$, $=""$ final score, $FS$, $=0$, collaborative investor interaction weight, $w_1$, collaborative investor feature weight, $w_2$ collaborative score threshold, $CBthresh$, collaborative score weight, $w_{CB}$, content score weight, $w_{CBS}$.
 \FOR {$i=[1, 2, 3, \hdots, n]$ $\&$ $j=[1, 2, 3, \hdots, m]$}
 \STATE \textbf{(1) Content Based Model:}
 \STATE Obtain all the companies in which investor ‘$i$’ has invested in from the training set, say, $comp \in [1, 2, \hdots, k]$.
 \FOR {$comp=[1, 2, \hdots, k]$}
 \STATE Obtain $compvec=$ vector representation for $comp$ and $jvec=$ vector representation for $j$.
 \STATE Calculate cosine similarity,\\ \hspace{12mm}$cos(j,comp) = \frac{jvec.compvec}{||jvec||.||compvec||}$
 \IF {$cos(j,comp) > CBS$}
 \STATE $CBS = cos(j,comp)$
 \STATE $CC = comp$ and $CCB =$ Extract the first sentence from description field
 \ENDIF 
 \ENDFOR
 \STATE \textbf{(2) Collaborative Based Model:}
 \STATE $u, s, d =$ Perform Singular Value Decomposition \cite{10.1007/BF02163027} on the ‘$sparse$’ interactive investor-company link matrix.\\
  ($u$ contains all investors with respect to their latent features)
 \STATE Obtain all the previous investors of company ‘$j$’ from the training set, say, $inv \in [1, 2, \hdots, l]$.
 \FOR {$inv=[1, 2, \hdots, l]$}
 \STATE Obtain $invvec=$ vector representation for $inv$ and $ivec=$ vector representation for $i$.
 \STATE Calculate cosine similarity, $sim1(i, inv) = \frac{u[i].u[inv]}{||u[i]||.||u[inv]||}$
 \STATE Calculate cosine similarity, $sim2(i, inv) = \frac{ivec.invvec}{||ivec||.||invvec||}$
 \STATE Calculate final similarity score,\\ \hspace{12mm}$sim(i,inv)=w_1*sim1+w_2*sim2$ 
 \IF {$sim(i,inv) > CB$}
 \STATE $CB = sim(i,inv)$
 \STATE $CI = inv$
 \ENDIF
 \ENDFOR
 \STATE \textbf{(3) Final Score:}
 \IF {$CB>CBthresh$}
 \STATE $FS(i,j)=w_{CBS}*CBS+w_{CB}*CB$
 \ELSE
 \STATE $FS(i,j)=CBS$
 \ENDIF
 \ENDFOR
 \end{algorithmic}
 \end{algorithm}

\subsection{Results}\label{results}

This section discusses the performance of the our proposed matching algorithm on a dataset containing 40,000 historical investor-company links. As discussed in section \ref{data}, 70\% of the historical investor-company links has been used as part of the training set and remaining 30\% as the test set. Also, it may be noted that the results shown are based on the threshold set at 0.75, i.e. any time the similarity score is greater than 0.75 we consider that a link exists between the said company and investor pair, and if similarity score is less than 0.75, no link exists. 

\begin{figure}[hb]
\begin{center}
\centerline{\includegraphics[width=\columnwidth]{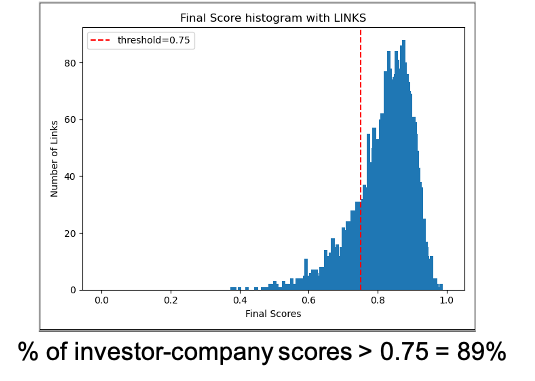}}
\vspace*{-5mm}
\caption{Histogram with historical links}
\label{fig:Histogram with historical links}
\end{center}
\vskip -0.2in
\end{figure}

Figure \ref{fig:Histogram with historical links} shows the histogram of the final similarity score obtained on our test set for which we know the ground-truth is that a link between said company and investor pair exists. As expected, in this case the histogram of similarity scores is skewed to the right and the model correctly predicts that there should be links between the investors and companies for 89\% of the test set.

\begin{figure}[ht!]
\begin{center}
\centerline{\includegraphics[width=\columnwidth]{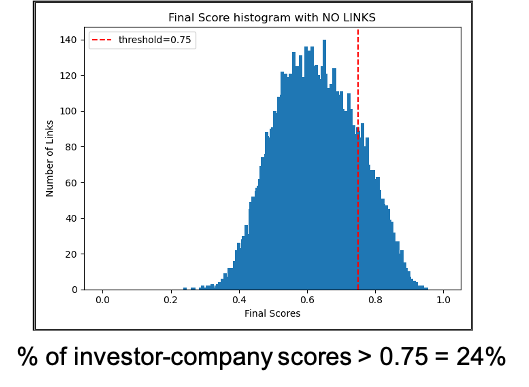}}
\vspace*{-5mm}
\caption{Histogram with currently no historical links}
\label{fig:Histogram with currently no historical links}
\end{center}
\vskip -0.2in
\end{figure}

Figure \ref{fig:Histogram with currently no historical links} shows a similar similarity-score histogram for a test set that contains randomly chosen company and investor pairs, and for which we know the ground truth that there is no historical link between the said company-investor pair. As expected, the histogram is skewed to the left and the model predicts that there should be links between the investors and companies only for 24\% of the test set. It may be noted that even an ideal model should have a small percentage of investor/company pairs for which it predicts that a link should exists, because these become the model's recommendations to investors/companies.

\begin{figure}[h!]
\begin{center}
\centerline{\includegraphics[width=\columnwidth]{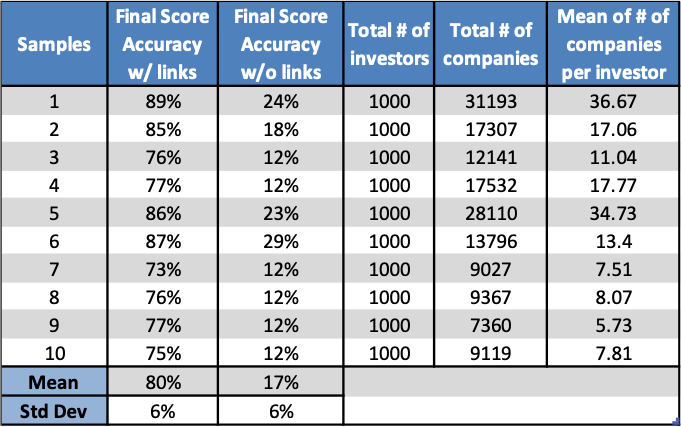}}
\caption{Stability Analysis over 10 independent samples}
\label{fig:Stability Analysis over 10 independent samples}
\end{center}
\vskip -0.2in
\end{figure}

Further, in order to test the stability of the matching algorithm, 10 independent samples were obtained from dataset, each containing 1000 investors. Figure \ref{fig:Stability Analysis over 10 independent samples} shows the performance of the model over these sampled dataset. The column ``Final Score Accuracy w/Links'' estimates the accuracy in correctly predicting a link when the ground-truth link exists. Also, the column, ``Final score Accuracy w/o links'' measures the percentage of companies for which we get a score above our threshold when we know that underlying data has no-links, i.e. historically no link existed between the said company and investor pair. We may re-iterate that these few percentage cases actually constitute the models recommendations, and while this percentage should be low, this should not be very close to zero, else we are not recommending anything. As can be observed the model has a good stability with a mean of 80\% performance and standard deviation of only 6\% for the test sets where there are existing historical links between the investors and companies. In case of the test set containing randomly chosen companies for which there is information that links do not exist between the investors and companies, the model is stable with a mean of 17\% performance and standard deviation of only 6\%.

\begin{figure}[h!]
\begin{center}
\centerline{\includegraphics[width=\columnwidth]{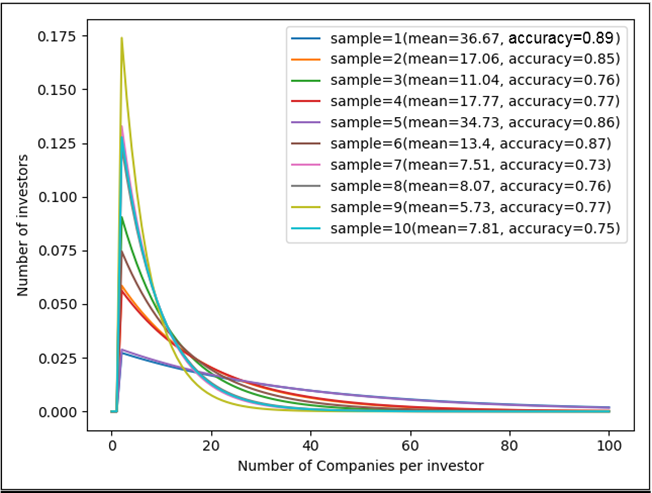}}
\caption{Stability Graph over 10 independent samples}
\label{fig:Stability Graph over 10 independent samples}
\end{center}
\vskip -0.3in
\end{figure}

To further analyze the standard deviation of 6\%, an exponential graph was constructed to observe how the performance of the model varies over these 10 independent samples with respect to the mean number of companies per investor. As can be observed in Figure \ref{fig:Stability Graph over 10 independent samples}, the more the mean number of companies per investor, the better the performance of the model is i.e. the more historical track records/preferences and investor-company interaction links we have, the better the model recommendations are.

Finally, in order to test the importance of the company features that are being used, we tested the model by taking one feature at a time. We also analyzed the model over a pair of features as well as triplets of features. The results are based on the same dataset of 40,000 historical investor-company links with a threshold set at 0.75. As can be observed in Table \ref{tab: Feature Importance Analysis} \footnote{To protect proprietary data, the feature details have been masked.}, features B and D of the company are the key driving features of the model. Also, individually, feature B brings noise in the results and feature D is very hard on the results. The combination of these two features brings the model performance very close to the performance of the model while using all the company features.

% \begin{figure}[h]
% \begin{center}
% \centerline{\includegraphics[width=\columnwidth]{figures/ExplainabilityFeatureImportance.png}}
% \caption{Explainability Analysis}
% \label{fig:Feature Importance Analysis}
% \end{center}
% \vskip -0.3in
% \end{figure}

\begin{table}[h]
\centering
\begin{adjustbox}{width=\columnwidth,center}
\begin{tabular}{|l|c|c|}
\hline
\multicolumn{1}{|c|}{\textbf{Features}} & \textbf{\begin{tabular}[c]{@{}c@{}}Final Score \\ Accuracy w/ \\ links\end{tabular}} & \textbf{\begin{tabular}[c]{@{}c@{}}Final Score \\ Accuracy \\ w/o links\end{tabular}} \\ \hline
\textbf{All}                            & \textbf{89\%}                                                                        & \textbf{24\%}                                                                         \\ \hline
Feature   A                             & 98\%                                                                                 & 80\%                                                                                  \\ \hline
\textbf{Feature   B}                             & \textbf{95\%}                                                                                 & \textbf{47\%}                                                                                 \\ \hline
Feature   C                             & 94\%                                                                                 & 60\%                                                                                  \\ \hline
\textbf{Feature   D}                            & \textbf{80\%}                                                                               & \textbf{18\%}                                                                                 \\ \hline
Feature   A, Feature B                  & 95\%                                                                                 & 47\%                                                                                  \\ \hline
Feature   A, Feature C                  & 94\%                                                                                 & 59\%                                                                                  \\ \hline
Feature   A, Feature D                  & 79\%                                                                                 & 17\%                                                                                  \\ \hline
Feature   B, Feature C                  & 95\%                                                                                 & 43\%                                                                                  \\ \hline
\textbf{Feature   B, Feature D}                  & \textbf{89\%}                                                                                 & \textbf{26\%}                                                                                  \\ \hline
Feature   C, Feature D                 & 79\%                                                                                 & 17\%                                                                                  \\ \hline
Feature   A, Feature B, Feature C       & 95\%                                                                                 & 43\%                                                                                  \\ \hline
\textbf{Feature   A, Feature B, Feature D}       & \textbf{89\%}                                                                                 & \textbf{25\%}                                                                                  \\ \hline
Feature   A, Feature C, Feature D       & 79\%                                                                                 & 17\%                                                                                  \\ \hline
\textbf{Feature   B, Feature C, Feature D}       & \textbf{88\%}                                                                                 & \textbf{24\%}                                                                                  \\ \hline
\end{tabular}
\end{adjustbox}
\vspace{0.1cm}
\caption{Explainability Analysis}
\label{tab: Feature Importance Analysis}
\end{table}

\vspace{-0.2in}

Finally, we also verified and confirmed that the investor-company matching algorithm is fair and is not biased based on management team's personal profile or location of company/investor.

\subsection{Human Analysis}\label{analysis}

In order to analyze the relevance of the recommendations produced by the model, we provided top 25 investor recommendations for four companies \footnote{Due to proprietary reasons, the company and investor names have been masked and fictitious names have been used.} including Software, LLC, Technology, LLC, Product, LLC and Marine, LLC to the subject matter experts (SMEs) for review. Out of these 100 investor-company matching pairs, 65 pairs were confirmed as relevant by the SMEs. Table \ref{tab: Human Analysis with High Scores} shows an example of the investor recommendation for each of the four companies for which the SMEs/humans gave a high score with their rationale. 

% \begin{figure}[ht]
% \begin{center}
% \centerline{\includegraphics[width=\columnwidth]{figures/Human_Pass.png}}
% \caption{Human Analysis with High Scores}
% \label{fig:Human Analysis with High Scores}
% \end{center}
% \vskip -0.2in
% \end{figure}

\begin{table}[h]
\centering
\begin{adjustbox}{width=\columnwidth,center}
\begin{tabular}{|c|c|c|c|c|}
\hline
\textbf{Company} & \textbf{\begin{tabular}[c]{@{}c@{}}Recommended\\ Investor\end{tabular}} & \textbf{\begin{tabular}[c]{@{}c@{}}Algo\\ Matching\\ Score\end{tabular}} & \textbf{\begin{tabular}[c]{@{}c@{}}Human \\ Score\end{tabular}} & \textbf{Human Rationale} \\ \hline \begin{tabular}[c]{@{}c@{}}Software,\\ LLC\end{tabular} & \begin{tabular}[c]{@{}c@{}}Software \\ Equity\end{tabular} & 94\% & HIGH                 
& \begin{tabular}[c]{@{}c@{}}Align with\\ industry focus\end{tabular}                        \\ \hline
\begin{tabular}[c]{@{}c@{}}Technology,\\ LLC\end{tabular} & \begin{tabular}[c]{@{}c@{}}Technology\\ Partners\end{tabular} & 90\%                         & HIGH & \begin{tabular}[c]{@{}c@{}}Early stage \\ investment track \\ record in media / \\ design software\end{tabular} \\ \hline Product, LLC & \begin{tabular}[c]{@{}c@{}}Product\\ Capital\end{tabular} & 90\% & HIGH                                                            & \begin{tabular}[c]{@{}c@{}}Aligned with\\ industry focus\end{tabular} \\ \hline
Marine, LLC & \begin{tabular}[c]{@{}c@{}}Marine\\ Partners\end{tabular} & 89\%                 & HIGH & \begin{tabular}[c]{@{}c@{}}Specific market\\ place vertical\end{tabular} \\ \hline
\end{tabular}
\end{adjustbox}
\vspace{0.1cm}
\caption{Human Analysis with High Scores}
\label{tab: Human Analysis with High Scores}
\end{table}

\vspace{-0.3in}

\noindent For 35 pairs, the SMEs/humans gave a low score and thought computer generated recommendations may not be relevant. Table \ref{tab: Human Analysis with Low Scores} shows an example of the investor recommendation for each of the four companies for which the SMEs/humans gave a low score along with their rationale for low score. Subsequently, we used our parameterized explanatory algorithm (described in next section) to algorithmically generate explanations, as to why the particular recommendation had been made. Following the explanations, the subject matter experts were convinced that the recommendations are indeed correct, and assigning them a low human score was actually due to limited breadth of human knowledge. This further illustrates the importance of mechanisms that enable explainability.

% \begin{figure}[h!]
% \begin{center}
% \centerline{\includegraphics[width=\columnwidth]{figures/Human_Fail.png}}
% \caption{Human Analysis with Low Scores}
% \label{fig:Human Analysis with Low Scores}
% \end{center}
% \vskip -0.3in
% \end{figure}

\begin{table}[h]
\centering
\begin{adjustbox}{width=\columnwidth,center}
\begin{tabular}{|c|c|c|c|c|}
\hline
\textbf{Company} & \textbf{\begin{tabular}[c]{@{}c@{}}Recommended\\ Investor\end{tabular}} & \textbf{\begin{tabular}[c]{@{}c@{}}Algo\\ Matching\\ Score\end{tabular}} & \textbf{\begin{tabular}[c]{@{}c@{}}Human \\ Score\end{tabular}} & \textbf{Human Rationale} \\ \hline \begin{tabular}[c]{@{}c@{}}Software,\\ LLC\end{tabular} & \begin{tabular}[c]{@{}c@{}}Software \\ Partners\end{tabular} & 94\% & LOW                 
& \begin{tabular}[c]{@{}c@{}}Not aligned with\\ industry focus\end{tabular}                        \\ \hline
\begin{tabular}[c]{@{}c@{}}Technology,\\ LLC\end{tabular} & \begin{tabular}[c]{@{}c@{}}Technology\\ Capital\end{tabular} & 90\% & LOW & \begin{tabular}[c]{@{}c@{}}Not aligned with\\ industry focus\end{tabular} \\ \hline Product, LLC & \begin{tabular}[c]{@{}c@{}}Product\\ Partners\end{tabular} & 90\% & LOW & \begin{tabular}[c]{@{}c@{}}Not aligned with\\ industry focus\end{tabular} \\ \hline
Marine, LLC & \begin{tabular}[c]{@{}c@{}}Marine\\ Equity\end{tabular} & 89\%                 & LOW & \begin{tabular}[c]{@{}c@{}}Not aligned with\\ industry focus\end{tabular} \\ \hline
\end{tabular}
\end{adjustbox}
\vspace{0.1cm}
\caption{Human Analysis with Low Scores}
\label{tab: Human Analysis with Low Scores}
\end{table}

\vspace{-0.3in}

\section{Parametrized Explanatory Algorithm}\label{algo}

As mentioned in Section \ref{intro}, the ability to explain why we make a particular recommendation is an essential requirement for our model, and is critical to ensure adoption of our model in real-life. The following sections describe the algorithm used to generate these explanations and provides some sample explanation examples produced using the same. It may be noted that the data required for generating these algorithmic explanations is directly taken from the intermediate results being generated by our matching algorithm.  

\vspace*{-0.1in}

\subsection{Algorithm}
% \vspace*{-6mm}
 \emph{Algorithm \ref{alg2}} describes our parameterized explanatory algorithm in a step-by-step manner. In our proposed algorithm, we have a predefined written template, and the parameters of the template are selected and filled using the intermediary results obtained from the matching algorithm. The explanation generated to explain the recommendation score between the investor $i$ and company $j$ is three-pronged:

\vspace*{-3mm}

\begin{algorithm}
  \caption{Parameterized Explanatory Algorithm}
  \label{alg2}
  \begin{algorithmic}[1]
  \STATE \textbf{Input 1:} For an investor $i$ where $i\in [1, 2, 3, \hdots, n]$ where $n$ is the number of investors in the dataset
 \STATE \textbf{Input 2:} For a company $j$ where $j\in [1, 2, 3, \hdots, m]$ where $m$ is the number of companies in the dataset
 \STATE \textbf{PREDEFINED EXPLANATION TEMPLATE: }\\
 The investor \it‘[param a]’ \rm is similar to \it‘[param b]’ \rm's previous investor \it‘[param c]’ \rm with score of \it‘[param d]’ \rm based on their industry focus preferences. Also, the investor \it‘[param a]’ \rm has invested in a company named \it‘[param e]’ \rm with a match score of \it‘[param f]’ \rm with \it‘[param b]’ \rm. \it‘[param e]’ \rm also \it‘[param g]’ \rm similar to \it‘[param b]’ \rm.
\FOR {$i=[1, 2, 3, \hdots, n]$ $\&$ $j=[1, 2, 3, \hdots, m]$}
\STATE Follow Steps 5-14 in Algorithm \ref{alg1} and obtain content based score, $CBS$, closest company, $CC$, and closest company business, $CCB$
\STATE Follow Steps 15-28 in Algorithm \ref{alg1} and obtain collaborative based score, $CB$, and closest investor, $CI$ 
\STATE SET $param\;a \leftarrow investor,\;i$
\STATE SET $param\;b \leftarrow company,\;j$
\STATE SET $param\;c \leftarrow closest\;investor,\;CI$
\STATE SET $param\;d \leftarrow collaborative\;based\;score,\;CB$
\STATE SET $param\;e \leftarrow closest\;company,\;CC$
\STATE SET $param\;f \leftarrow content\;based\;score,\;CBS$
\STATE SET $param\;g \leftarrow closest\;company\;business,\;CCB$
\ENDFOR
  \end{algorithmic}
\end{algorithm}

% \begin{figure*}[h!]
% \begin{center}
% \centerline{\includegraphics[width=1.0\textwidth]{figures/ExplainabilityHumanFailure.png}}
% \caption{Parametrized Explanatory Algorithm Results}
% \label{fig:Parametrized Explanatory Algorithm Results}
% \end{center}
% \vskip -0.3in
% \end{figure*}

\vspace{-0.6mm}

\begin{itemize}
    \item \textbf{Investor-Investor Similarity:} This part clarifies that the investor $i$ is most similar to which of the previous investors of company $j$ and with how much score.
    \item \textbf{Company-Company Similarity:} This part clarifies that the company $j$ is most similar to which of the investor $i$'s portfolio of existing invested companies and with how much score.
    \item \textbf{Company Description:} Finally, this part shows the key business area that the company, which is the most similar to company $j$ (as obtained in the above part), works on. This company description gives a more detailed explanation at the feature level as to why we are recommending a link between the investor $i$ and company $j$.
\end{itemize}

\begin{table*}[h!]
\centering
\begin{adjustbox}{width=1.0\textwidth,center}
\begin{tabular}{|c|c|c|c|l|l|l|}
\hline
\textbf{Company} & \textbf{\begin{tabular}[c]{@{}c@{}}Recommended   \\ Investor\end{tabular}} & \textbf{\begin{tabular}[c]{@{}c@{}}Algo   \\ Matching \\ Score\end{tabular}} & \textbf{\begin{tabular}[c]{@{}c@{}}Human   \\ Score\end{tabular}} & \multicolumn{1}{c|}{\textbf{\begin{tabular}[c]{@{}c@{}}Human   \\ Rationale\end{tabular}}} & \multicolumn{1}{c|}{\textbf{\begin{tabular}[c]{@{}c@{}}Declared Target \\ Industries for Investor\end{tabular}}}                                                                   & \multicolumn{1}{c|}{\textbf{Algo Explainability Rationale}}                                                                                                                      \\ \hline
\begin{tabular}[c]{@{}c@{}}Software,   \\ LLC\end{tabular}   & \begin{tabular}[c]{@{}c@{}}Software   \\ Partners\end{tabular}             & 92\%                                   & LOW & \begin{tabular}[c]{@{}l@{}}Not aligned \\ with industry \\ focus\end{tabular}              & Software                                                                      & \begin{tabular}[c]{@{}l@{}}The investor "\underline{Software Partners}" is similar to \\ "\underline{Software, LLC}"'s previous investor "\underline{Software Capital}" \\ with score of "\underline{0.92}" based on their industry focus \\ preferences. Also, the investor "\underline{Software Partners}" \\ has invested in a company named "\underline{Platform, LLC}" \\ with a match score of "\underline{0.93}" with "\underline{Software, LLC}". \\ "\underline{Platform, LLC}" also "\underline{develops mobility technology} \\ \underline{platform}" similar to "\underline{Software, LLC}".\end{tabular}      \\ \hline
\begin{tabular}[c]{@{}c@{}}Technology,   \\ LLC\end{tabular} & \begin{tabular}[c]{@{}c@{}}Technology   \\ Capital\end{tabular}            & 89\%                                   & LOW & \begin{tabular}[c]{@{}l@{}}Not aligned \\ with industry \\ focus\end{tabular} & \begin{tabular}[c]{@{}l@{}}Biotechnology,   \\ Commercial Products \\ and Services, \\ Communications and \\ Networking, Energy   \\ Equipment, \\ Infrastructure, \\ Transportation\end{tabular}                     & \begin{tabular}[c]{@{}l@{}}The investor "\underline{Technology Capital}" is similar to \\ "\underline{Technology, LLC}"'s previous investor "\underline{Technology} \\ \underline{Investments}" with score of "\underline{0.91}" based on their \\ industry focus preferences. Also, the investor \\ "\underline{Technology Capital}" has invested in a company \\ named "\underline{Mobile, LLC}" with a match score of "\underline{0.82}" \\ with "\underline{Technology, LLC}". "\underline{Mobile, LLC}" also "\underline{develops} \\ \underline{web/mobile applications}" similar to "\underline{Technology, LLC}".\end{tabular} \\ \hline
\begin{tabular}[c]{@{}c@{}}Product,   \\ LLC\end{tabular}    & \begin{tabular}[c]{@{}c@{}}Product   \\ Partners\end{tabular}              & 92\%                                   & LOW  & \begin{tabular}[c]{@{}l@{}}Not aligned \\ with industry \\ focus\end{tabular}    & \begin{tabular}[c]{@{}l@{}}Commercial Products, \\ Commercial Services, \\ Communications and \\ Networking, Healthcare \\ Services, IT Services, \\ Software\end{tabular}    & \begin{tabular}[c]{@{}l@{}}The investor "\underline{Product Partners}" is similar to "\underline{Product,} \\ \underline{LLC}"'s previous investor "\underline{Product Ventures}" with \\ score of "\underline{0.92}" based on their industry focus \\ preferences. Also, the investor "\underline{Product Partners}" has\\  invested in a company named "\underline{Market, LLC}" with a \\ match score of "\underline{0.89}" with "\underline{Product, LLC}". "\underline{Market,} \\ \underline{LLC}" also "\underline{develops online platform to sell personal} \\ \underline{products}" similar to "\underline{Product, LLC}".\end{tabular}  \\ \hline
\begin{tabular}[c]{@{}c@{}}Marine,   \\ LLC\end{tabular}     & \begin{tabular}[c]{@{}c@{}}Marine   \\ Equity\end{tabular}                 & 89\%                                   & LOW & \begin{tabular}[c]{@{}l@{}}Not aligned \\ with industry \\ focus\end{tabular}              & \begin{tabular}[c]{@{}l@{}}Communications, \\ Distributors/Wholesale, \\ Industrial Supplies, \\ Media and Information \\ Services, Metals and \\ Minerals, Restaurants, \\ Hotels, Leisure, \\ Software\end{tabular} & \begin{tabular}[c]{@{}l@{}}The investor "\underline{Marine Equity}" is similar to "\underline{Marine,} \\ \underline{LLC}"'s previous investor "\underline{Marine Capital}" with score \\ of "\underline{0.85}" based on their industry focus preferences. \\ Also, the investor "\underline{Marine Equity}" has invested in a \\ company named "\underline{Travel, LLC}" with a match score of \\ "\underline{0.89}" with "\underline{Marine, LLC}". "\underline{Travel, LLC}" also "\underline{develops} \\ \underline{online marine platform}" similar to "\underline{Marine, LLC}".\end{tabular}   \\ \hline
\end{tabular}
\end{adjustbox}
\vspace{0.1cm}
\caption{Parametrized Explanatory Algorithm Results}
\vspace{-0.2in}
\label{tab: Parametrized Explanatory Algorithm Results}
\end{table*}

% \vspace{-0.3in}
\vspace{-0.1in}

\subsection{Results}

Table \ref{tab: Parametrized Explanatory Algorithm Results} \footnote{Due to proprietary reasons, the company and investor names have been masked and fictitious names have been used.} shows the same examples as were shown in Table \ref{tab: Human Analysis with Low Scores} along with the explanations generated. The column 'Algo Explainability Rationale' shows the explanations that were generated algorithmically with parameters filled (highlighted as underlined text in Table \ref{tab: Parametrized Explanatory Algorithm Results}) into the predefined explanation template. For instance, in order to explain the high recommendation score between the company $Software,\:LLC$ and investor $Software\;Partners$, we selected and filled the three-tonged explanations as follows: (a) Investor-Investor Similarity: This part clarifies that investor $Software\;Partners$ is similar to the the company $Software,\:LLC$'s previous investor $Software\;Capital$ with a score of 0.92. (b) Company-Company Similarity: This part clarifies that the company $Software,\:LLC$ is most similar to the company $Platform,\:LLC$ with a score of 0.93. $Platform,\:LLC$ is one of the companies in which $Software\;Partners$ has previously invested in. (c) Company Description: Finally, this part clarifies that both $Software,\:LLC$ and $Platform,\:LLC$ develop mobility technology platforms.

The investor-matching pairs that were provided to the subject matter experts (SMEs) for review, as detailed in the Human Analysis section, were again submitted with these explanations for review. The SMEs accepted the explanations and were convinced that all the recommendations that were provided by the algorithm are relevant.

\section{Conclusion \& Future Work}\label{conclusion}

In this paper, we proposed a novel representation learning based recommendation system that works very well with extremely small datasets. We also highlighted the importance of explainability in recommendation engines, and demonstrated how we can couple recommendation engine with a parameterized explanation generation algorithm to build an explainable recommendation engine. We demonstrated the ability of our algorithm to work very well, even with limited dataset, on company-investor matching problem, a problem which has long been considered a highly specialized human cognitive decision making process. Further, we also demonstrated how explainable recommendations helped with real-life adoption of our system. We also verified and confirmed that the algorithm is fair and is not biased based on management team's personal profile or location of company/investor.

There are various avenue to further improve and build upon this work. For instance, the similarity estimation between companies and investors can be improved by using graph-based approaches. Further, we could further  improve company/investor vector representations by leveraging more features such as company/investor valuations, news articles, etc. Additionally, our parameterized explanatory engine can be further improved by doing more a granular analysis and provide explanations regarding which exactly which feature impacts the recommendation  score the most. 

\paragraph{\textbf{Disclaimer}}
This paper was prepared for informational purposes by
the Artificial Intelligence Research group of JPMorgan Chase \& Co\. and its affiliates (``JP Morgan''),
and is not a product of the Research Department of JP Morgan.
JP Morgan makes no representation and warranty whatsoever and disclaims all liability,
for the completeness, accuracy or reliability of the information contained herein.
This document is not intended as investment research or investment advice, or a recommendation,
offer or solicitation for the purchase or sale of any security, financial instrument, financial product or service,
or to be used in any way for evaluating the merits of participating in any transaction,
and shall not constitute a solicitation under any jurisdiction or to any person,
if such solicitation under such jurisdiction or to such person would be unlawful.
\\
© 2021 JPMorgan Chase \& Co. All rights reserved.

\bibliographystyle{ACM-Reference-Format}
\bibliography{Parametrized_Explanations_for_Investor_to_Company_Matching}

\end{document}